\pdfoutput=1
\documentclass[twoside,twocolumn,a4paper]{article}

\usepackage{blindtext} 

\usepackage[sc]{mathpazo} 
\usepackage[T1]{fontenc} 
\usepackage{microtype} 

\usepackage[english]{babel} 

\usepackage[left=25mm, right=20mm,top=32mm,columnsep=20pt]{geometry} 
\usepackage[hang, small,labelfont=bf,up,textfont=it,up]{caption} 
\usepackage{booktabs} 

\usepackage{lettrine} 

\usepackage{enumitem} 
\setlist[itemize]{noitemsep} 

\usepackage{abstract} 

\usepackage{titlesec} 
\renewcommand\thesection{\Roman{section}} 
\renewcommand\thesubsection{\roman{subsection}} 
\titleformat{\section}[block]{\large\scshape\centering}{\thesection.}{1em}{} 
\titleformat{\subsection}[block]{\large}{\thesubsection.}{1em}{} 

\usepackage{fancyhdr} 
\usepackage{titling} 

\usepackage{hyperref} 

\usepackage{upgreek} 
\usepackage{graphicx} 
\usepackage{graphbox} 
\usepackage{stfloats} 
\usepackage{lineno}
\usepackage{xcolor}

\renewcommand{\neq}[1]{$\cdot$10$^{#1}$\,n$_\textrm{eq}$/cm$^2$}

\pretitle{\begin{center}\Huge\bfseries} 
\posttitle{\end{center}} 
\title{ATLAS ITk Pixel Detector Overview} 
\author{%
\textsc{Lingxin Meng\thanks{Corresponding author}, on behalf of the ATLAS Collaboration} \\[1ex] 
\normalsize CERN \\ 
\normalsize \href{mailto:lingxin.meng@cern.ch}{lingxin.meng@cern.ch} 
}
\date{Talk presented at the International Workshop on Future Linear Colliders (LCWS2021), 15-18 March 2021. C21-03-15.1.} 
\renewcommand{\maketitlehookd}{%
\begin{abstract}
\noindent 
For the High Luminosity upgrade of the Large Hadron Collider the current ATLAS Inner Detector will be replaced by an all-silicon Inner Tracker. The pixel detector will consist of five barrel layers and a number of rings, resulting in about 13\,m$^2$ of instrumented area. Due to the huge non-ionising fluence (1\neq{16}) and ionising dose (5\,MGy), the two innermost layers, instrumented with 3D pixel sensors and 100\,$\upmu$m thin planar sensors, will be replaced after about five years of operation. Each pixel layer comprises hybrid detector modules that will be read out by novel ASICs, implemented in 65\,nm CMOS technology, with a bandwidth of up to 5\,Gbit/s. Data will be transmitted optically to the off-detector readout system. To save material in the servicing cables, serial powering is employed for the supply voltage of the readout ASICs. Large scale prototyping programmes are being carried out by all subsystems.

\noindent This paper will give an overview of the layout and current status of the development of the ITk Pixel Detector.

\end{abstract}
}

\begin{document}
\maketitle
\makeatletter{\renewcommand*{\@makefnmark}{}
\footnotetext{\\\color{darkgray}Copyright 2021 CERN for the benefit of the ATLAS Collaboration. Reproduction of this article or parts of it is allowed as specified in the CC-BY-4.0 license.}\makeatother}

\section{Introduction}

\begin{figure}[t]
\includegraphics[width=0.48\textwidth]{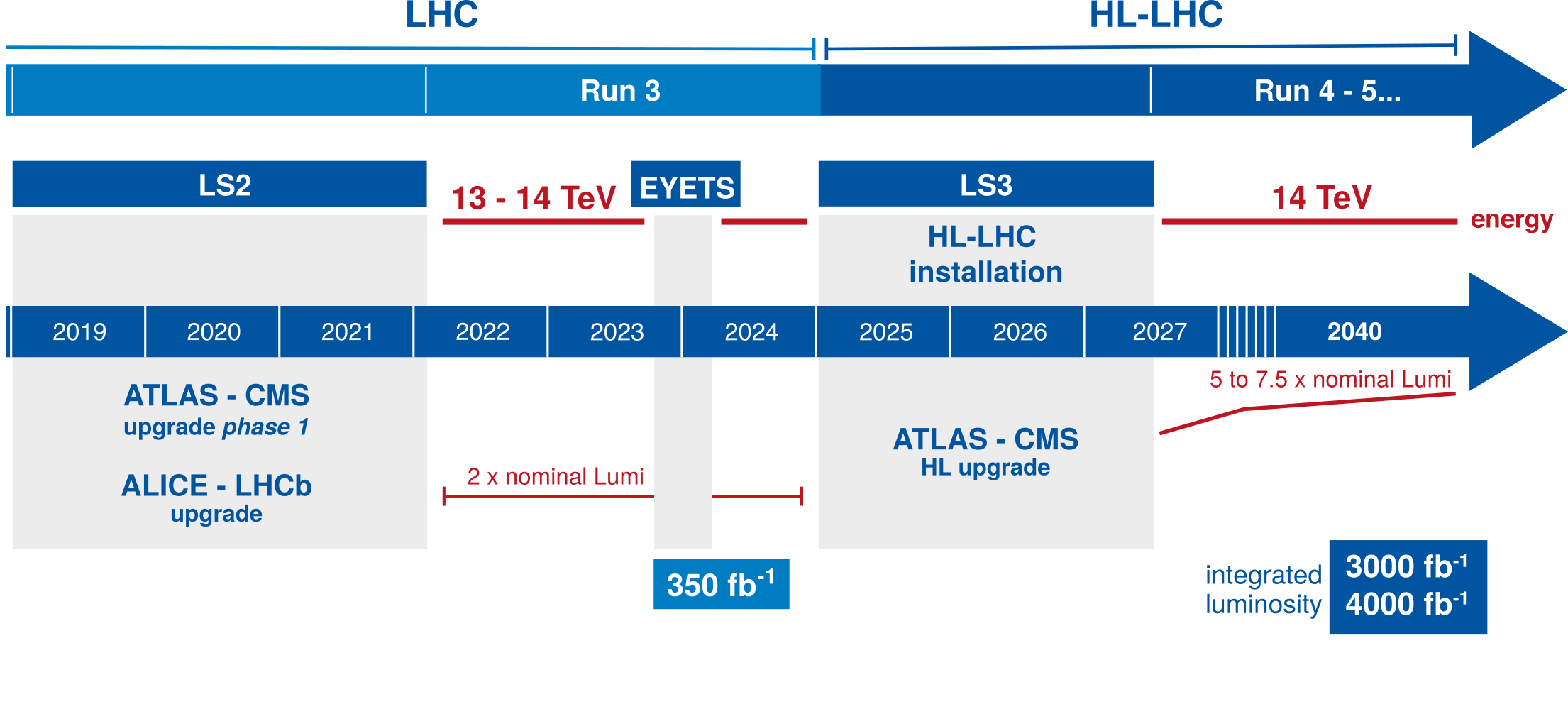}
\caption{Timeline of the Large Hadron Collider in the coming years. The High Luminosity upgrade is planned for Long Shutdown (LS) 3 \cite{www:hl-schedule}.}
\label{fig:hl-lhc}
\end{figure}

\lettrine[nindent=0em,lines=2]{I}n the High Luminosity (HL) era of the Large Hadron Collider (LHC), the average number of collisions per bunch crossing (BC) will increase from the current 30 to about 200, resulting in an increase of the peak luminosity from 1$\cdot 10^{34}$\,cm$^{-2}$s$^{-1}$ to 5--7$\cdot 10^{34}$\,cm$^{-2}$s$^{-1}$. The integrated luminosity of 350\,fb$^{-1}$ at the end of Run 3 will accumulate more rapidly up to 4000\,fb$^{-1}$ until the end of the HL programme. The requirement for the pixel detector of the ATLAS experiment \cite{atlas} and its readout also becomes more demanding: the occupancy is kept below 1\% by increasing the detector granularity with the frontend chip operating at 1\,MHz trigger rate. The innermost pixel system has to withstand a harsher radiation envirionment up to 2\neq{16}. This HL upgrade is planned to take place during the Long Shutdown (LS) 3 as indicated in Figure \ref{fig:hl-lhc}.

\section{ATLAS Inner Tracker Upgrade}
\begin{figure}[h]
\includegraphics[width=0.48\textwidth]{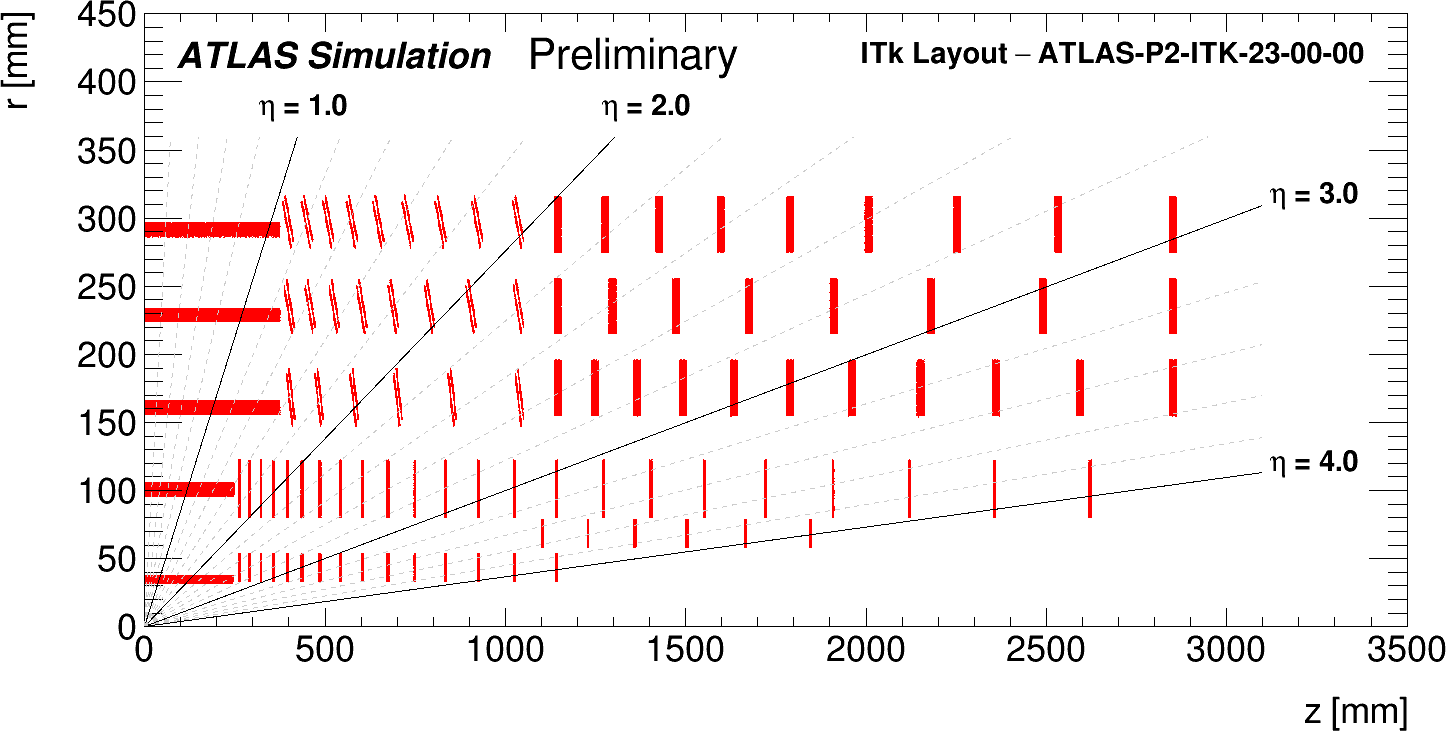}
\caption{Layout schematic of active elements in the r-z projection of the ITk Pixel detector \cite{itk-layout}.}
\label{fig:itkpixellayout}
\end{figure}

\begin{figure}
{\includegraphics[width=0.24\textwidth]{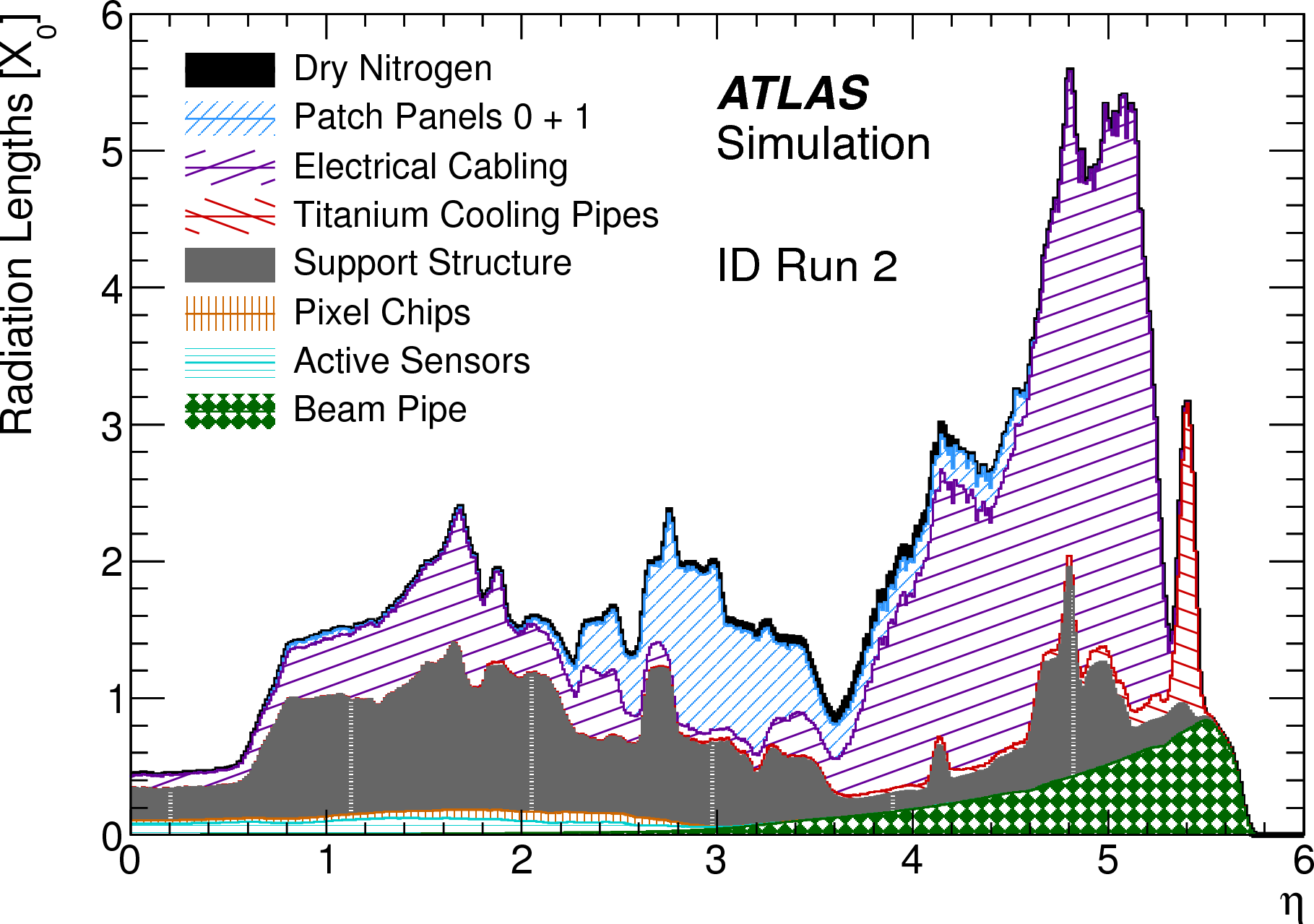}\hspace{0.005\textwidth}\includegraphics[width=0.23\textwidth]{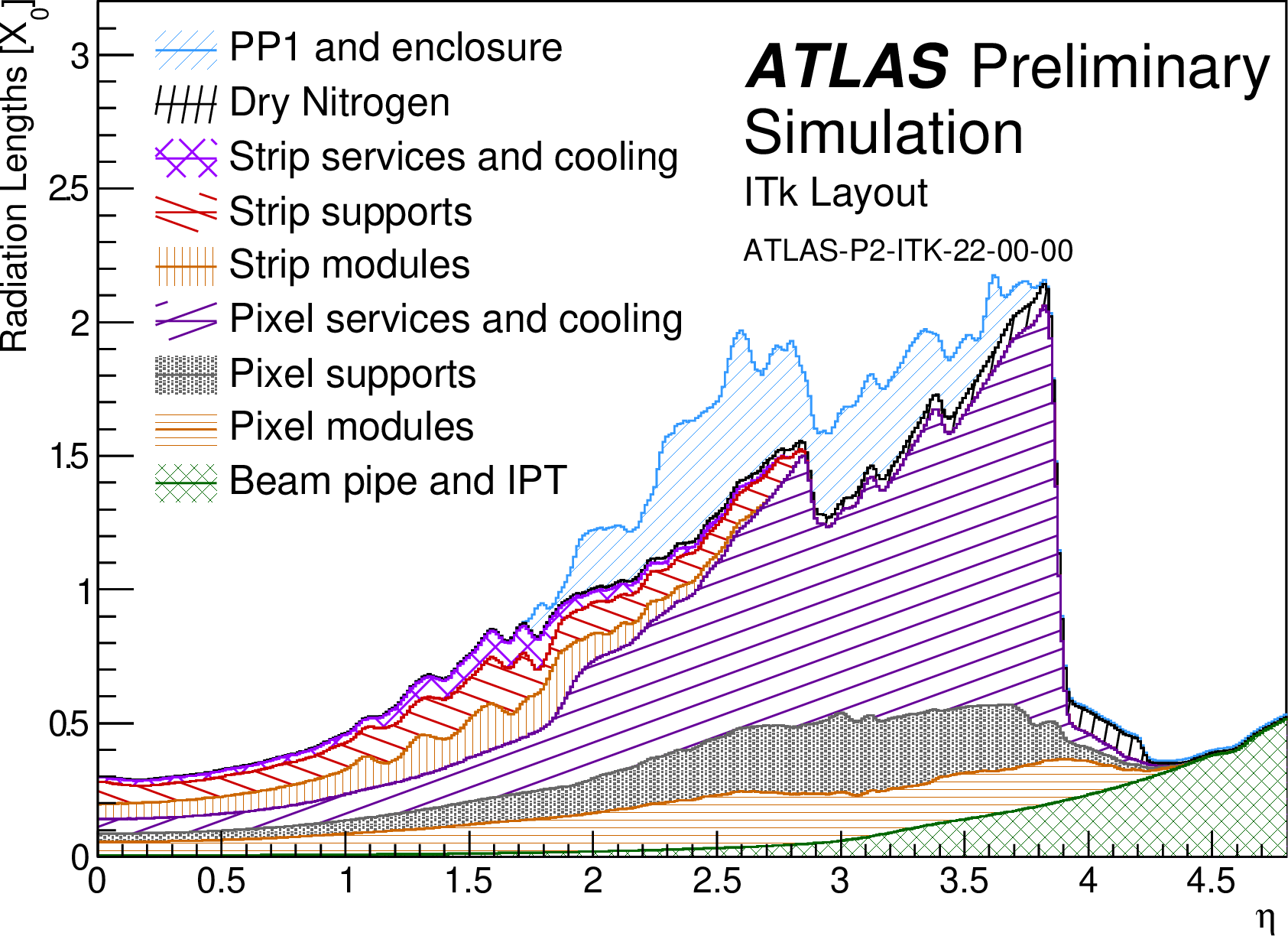}}
\caption{Comparison of the material budget in the detector between the current Inner Detector during Run 2 \cite{itk-pixel-tdr} and ITk \cite{itk-materialbudget}.}
\label{fig:materialbudget}
\end{figure}

For the HL upgrade of the ATLAS detector, the current Inner Detector (ID), consisting of silicon pixel, silicon strip detectors and Transition Radiation Tracker (TRT) is going to be replaced by an all-silicon Inner Tracker (ITk) \cite{itk-pixel-tdr, itk-strip-tdr}, which has  13\,m$^2$ of pixel detectors with 5 billion readout channels and 160\,m$^2$ of strip detectors with 50 million readout channels. The layout of active elements in the ITk Pixel detector in the r-z projection is shown in Figure \ref{fig:itkpixellayout}. The five-layer design provides a higher tracking coverage of up to $|\eta|=4$, while at the same time multiple scattering is minimised by greatly reducing the material budget compared to the current ID, as shown in Figure \ref{fig:materialbudget}.

Sections \ref{sec:sensors}, \ref{sec:chip} and \ref{sec:module} will introduce the sensors, readout chips and pixel modules as building blocks of the pixel detector, with which larger structures are built and described in sections \ref{sec:serialpowering} to \ref{sec:systemtest}. 

\subsection{Sensors}
\label{sec:sensors}
The pixel sensors are required to have an efficiency of $>$98.5\% before, and $>$97\% after an irradiation with a corresponding end-of-life fluence \cite{itk-pixel-tdr}. Two technologies are used for the ITk Pixel detector: 3D sensors in the innermost layer (L0) in triplet modules, consisting of three single frontend-sensor assemblies on a triplet flex, and planar sensors in quad modules in all other layers (L1--L4), consisting of four frontend chips on a quad sensor tile on a quad flex. Both technologies use n-implant in p-substrate with a generally thinner active thickness compared to the current pixel detector, thus charge and hit efficiencies are saturated at lower bias voltages which lead to a reduction of dissipated power. At the moment, a market survey is finished for both technologies and pre-production is ongoing.


\subsubsection{3D Sensors}
\begin{figure}
\centering
\includegraphics[width=0.40\textwidth]{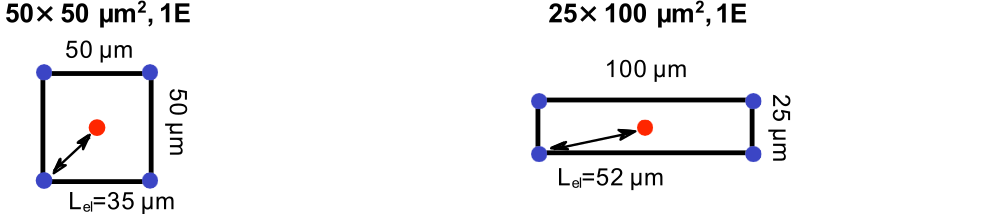}
\includegraphics[width=0.48\textwidth]{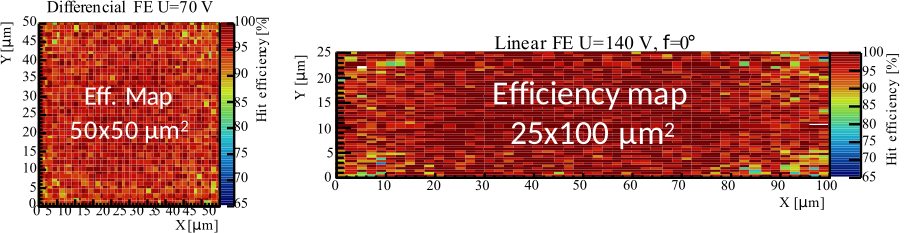}
\caption{Top: pixel sizes used in 3D sensors with a single collection electrode (red) \cite{gemmeVertex2020}. Bottom: Efficiency maps for different pixel sizes \cite{gemmeVertex2020}.}
\label{fig:3defficiency}
\end{figure}

3D sensors with a pixel size of 50$\times$50\,$\upmu$m$^2$ and 25$\times$100\,$\upmu$m$^2$ with a single collection electrode in the centre and a good efficiency, as shown in Figure \ref{fig:3defficiency}, are used in the endcap and barrel regions of L0, respectively. They consist of 150\,$\upmu$m active thickness plus 100\,$\upmu$m support wafer. 3D sensors are intrinsically more radiation tolerant and therefore used in the innermost layer which is only 34\,mm from the beam. It will be replaced after 2000\,fb$^{-1}$ with an expected fluence of 1.3\neq{16} as part of the Inner pixel system replacement \cite{itk-pixel-tdr}.

\subsubsection{Planar Sensors}
\begin{figure}
\centering
\includegraphics[align=c,height=27mm]{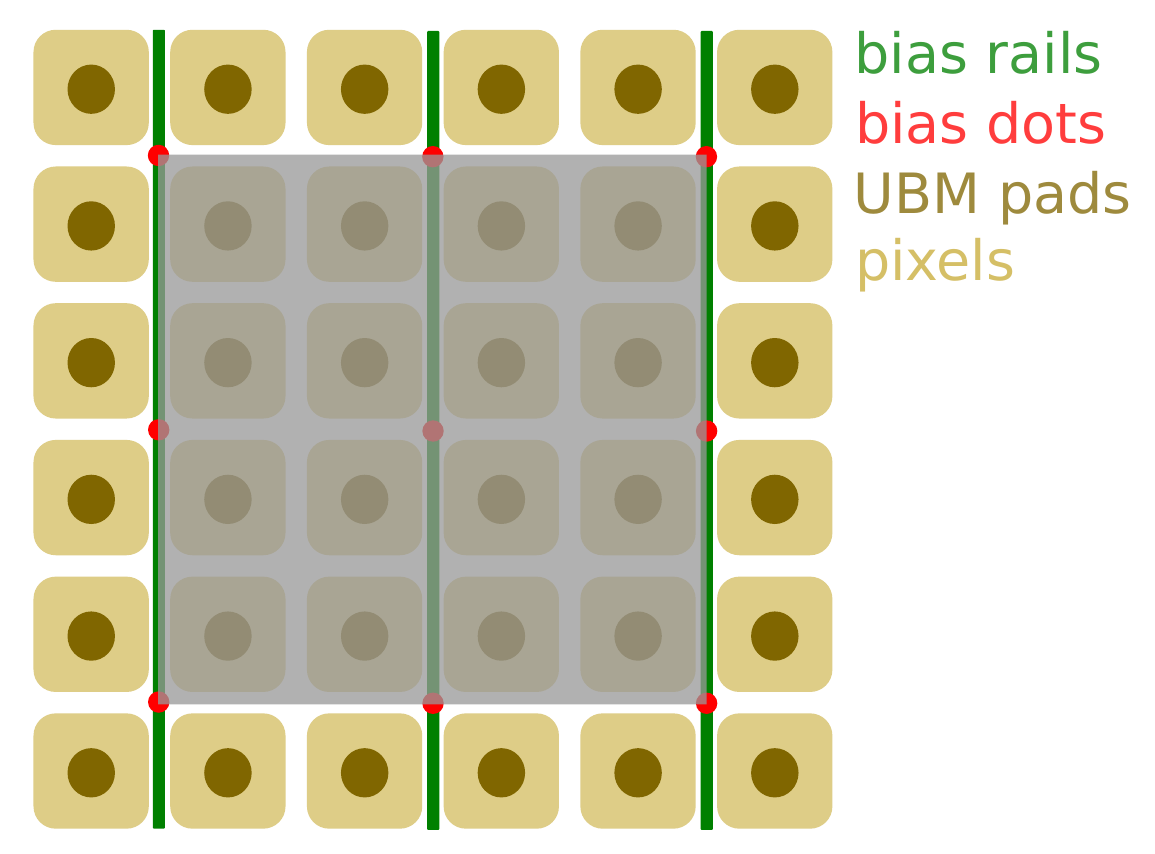}
\includegraphics[align=c,height=30mm]{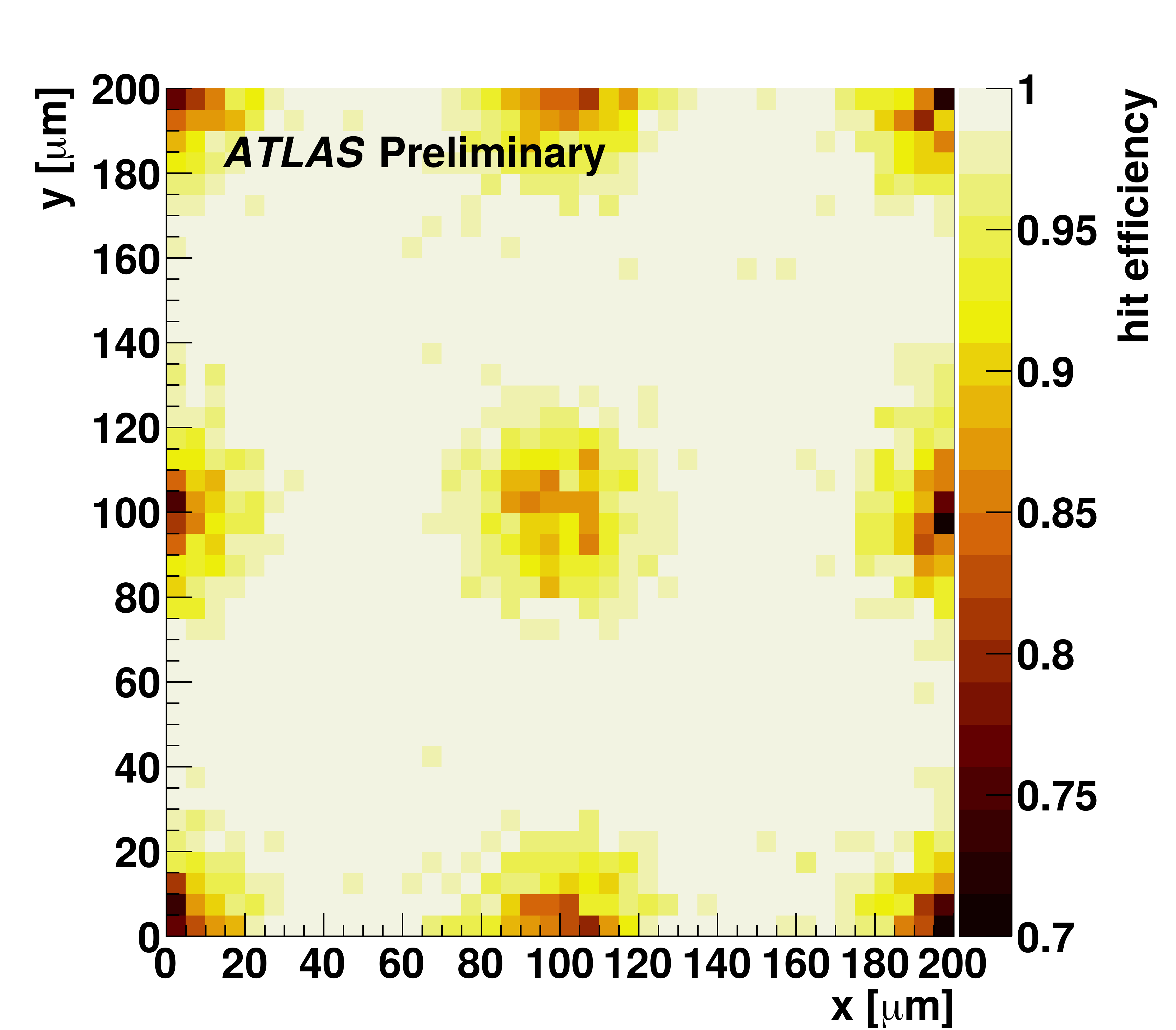}

\caption{Pixel matrix design with punch-through bias structure (left) and efficiency map (right) of $4\times4$ pixels (grey area) of a planar sensor before irradiation \cite{planarpixelMS}.}
\label{fig:planarefficiency}
\end{figure}

Planar sensors have a pixel size of 50$\times$50\,$\upmu$m$^2$ in all regions. The thickness is 100\,$\upmu$m in L1 in the Inner System and 150\,$\upmu$m in the outer layers. There are designs with punch-through bias structure as shown in Figure \ref{fig:planarefficiency} left and without (not shown). The efficiency of a $4\times4$ pixel region (grey area) before irradiation is shown in Figure \ref{fig:planarefficiency} right, where a slightly lower efficiency is observed in regions with punch-through bias dots \cite{planarpixelMS}.


\subsection{RD53 Readout Chip}
\label{sec:chip}
\begin{figure}
\includegraphics[width=0.48\textwidth]{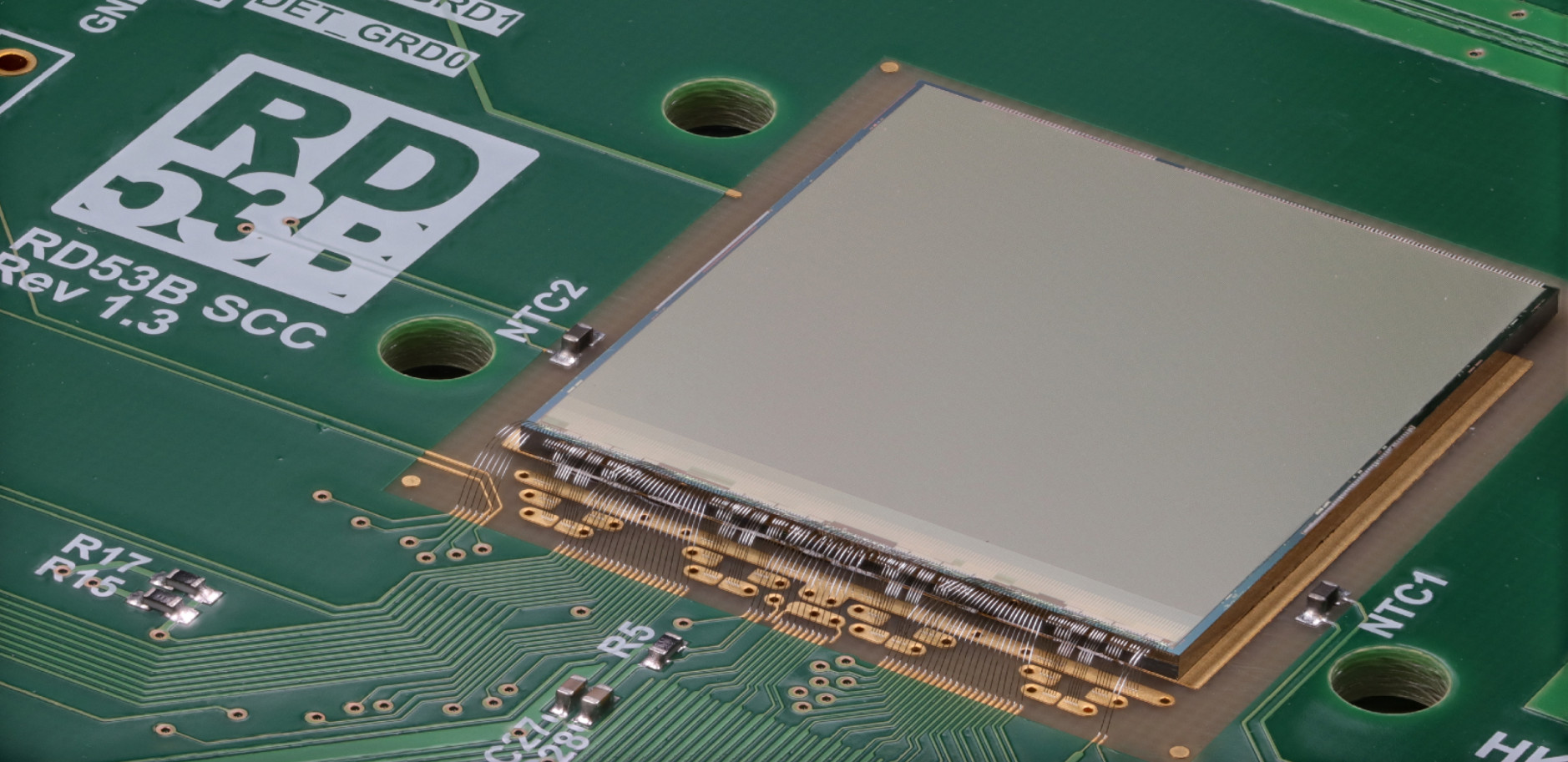}
\caption{An ITkPixV1 chip on a single chip card \cite{timonVertex2020}.}
\label{fig:itkpixv1scc}
\end{figure}

The frontend (FE) readout chip is developed by the RD53 collaboration \cite{www:rd53}, consisting of 24 institutes from Europe and the United States, as a common readout chip for ATLAS and CMS pixel detectors. It evolved from the RD53A prototype \cite{rd53amanual}, consisting of three different frontend flavours, into ITkPix \cite{rd53bmanual} for ATLAS with 400$\times$384 pixels using differential frontend, and CROC for CMS wtih 432$\times$336 pixels using linear frontend. ITkPixV1, shown in Figure \ref{fig:itkpixv1scc} on a single chip card (SCC), was submitted in March 2020 and is currently being verified. Submission of the final version, ITkPixV2, is foreseen for the end of 2021.

\begin{figure}[h]
\includegraphics[width=0.48\textwidth]{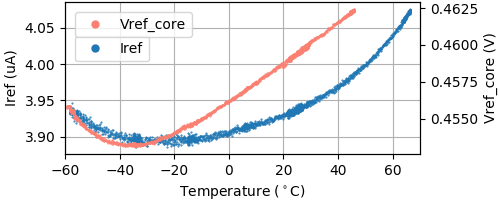}
\caption{Main reference current and voltage of the ITkPixV1 show only a change of less than 4\% over 120\,$^\circ$C temperature range \cite{timonVertex2020}.}
\label{fig:temperature}
\end{figure}

\begin{figure}[h]
\includegraphics[width=0.48\textwidth]{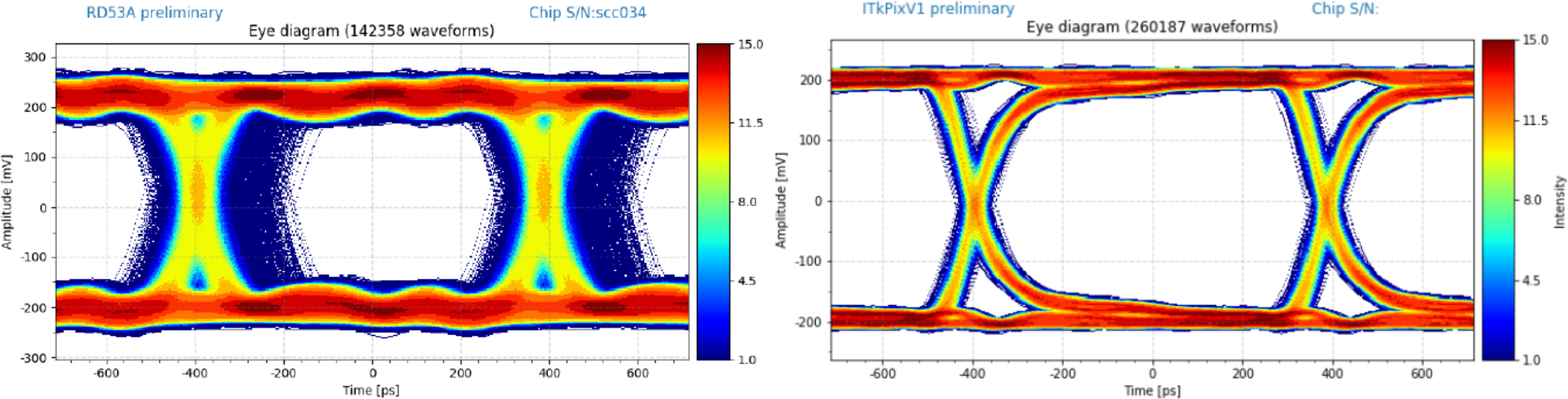}
\caption{Eye diagrammes of command data recovery (CDR) and PLL from RD53A (left) to ITkPixV1 (right). The jitter improved from 29.94\,ps RMS to 11.2\,ps RMS, respectively \cite{loddoTrento2021}.} 
\label{fig:jitter}
\end{figure}

\begin{figure}[b]
\includegraphics[width=0.48\textwidth]{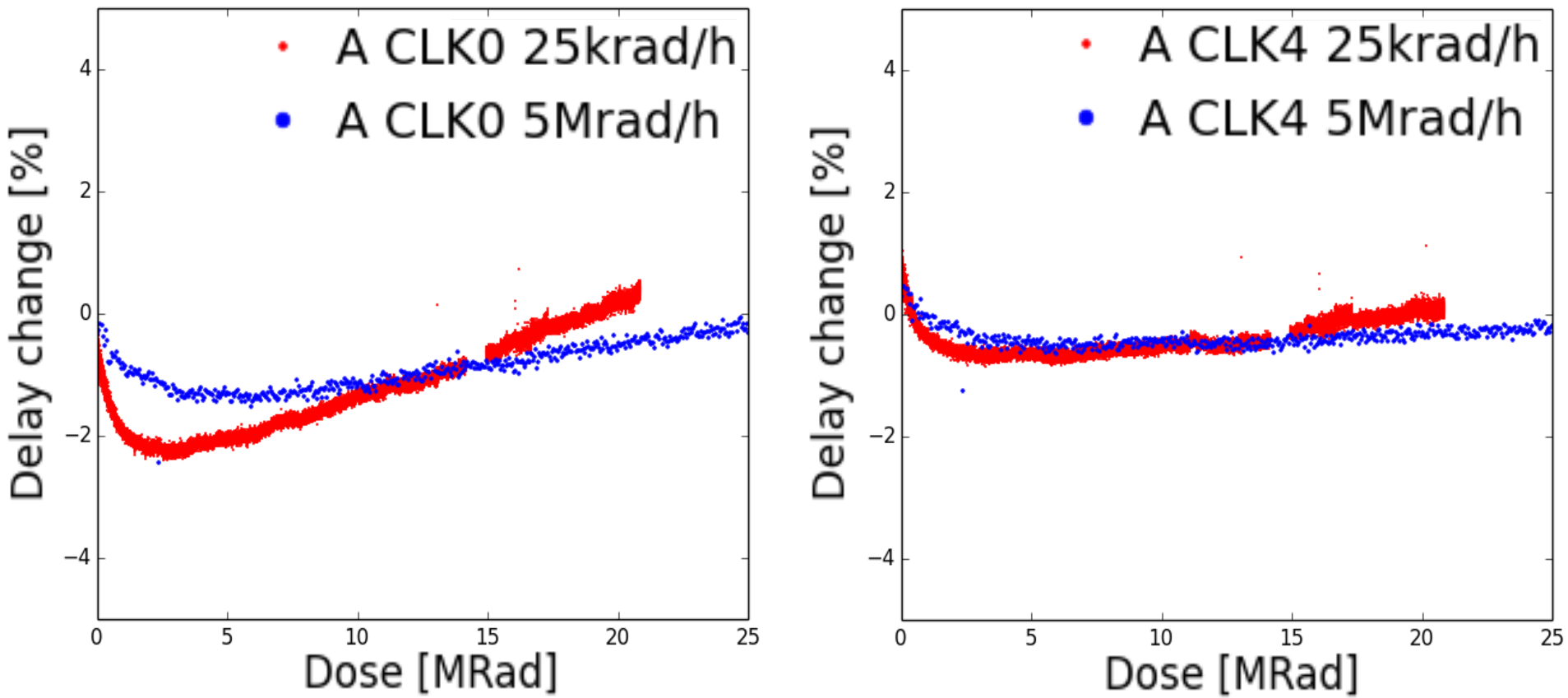}
\caption{X-ray irradiations comparing high (blue) and low (red) dose rate on ring oscillators of lengths 0 (left) and 4 (right) on the ITkPixV1 \cite{loddoTrento2021}.}
\label{fig:xray}
\end{figure}

Apart from an issue with the time-over-threshold (ToT) latch which is fixed in version 1.1, Figures \ref{fig:temperature}, \ref{fig:jitter} and \ref{fig:xray} show a small selection of verifications tests. The chips main reference current and voltage are stable within 4\% over a large temperature range of up to 120\,$^\circ$C (Figure \ref{fig:temperature}. The eye-diagramme of command data recovery (CDR) and PLL in Figure \ref{fig:jitter} shows an improved jitter from RD53A with 29.94\,ps RMS to ITkPixV1 with 11.2\,ps RMS \cite{loddoTrento2021}. X-ray irradiation campaignes are ongoing with ring oscillators as test structures. Figure \ref{fig:xray} compares high dose rate of 5\,Mrad/h (blue) with a low dose rate of 25\,krad/h (red) on the CLK gates of lengths 0 and 4 \cite{loddoTrento2021}. While it is faster to accumulate ionising dose at end-of-life with a high rate, low rate of ionising radiation that reflects realistic irradiation conditions in the detector generates more degradation at the same accumulated dose, which can be seen from the steeper slope. Studying both rates could allow an estimate of scalability of the degradation.

\begin{table}
\caption{Technical specifications of ITkPix \cite{itk-pixel-tdr,rd53bmanual}.}
\centering
\begin{tabular}{|l|l|}
\hline
Chip size		& $20\times 21\,$mm$^2$\\
\hline
Pixel size		& $50\times50\,\upmu$m$^2$\\
\hline
Hit rate		& 3\,GHz/cm$^2$\\
\hline
Trigger rate	& 1\,MHz\\
\hline
Latency			& 12.8\,$\upmu$s \\
\hline
Data rate		& 5.12\,Gbit/s\\
\hline
Low threshold	& 600\,e\\
\hline
Radiation tolerance	& 500\,Mrad\\
\hline
Low power		& 4\,$\upmu$A/pixel\\
\hline
\end{tabular}
\label{tab:itkpixspecs}
\end{table}


\subsection{Pixel Module}
\label{sec:module}
\begin{figure}[t]
\includegraphics[width=0.48\textwidth]{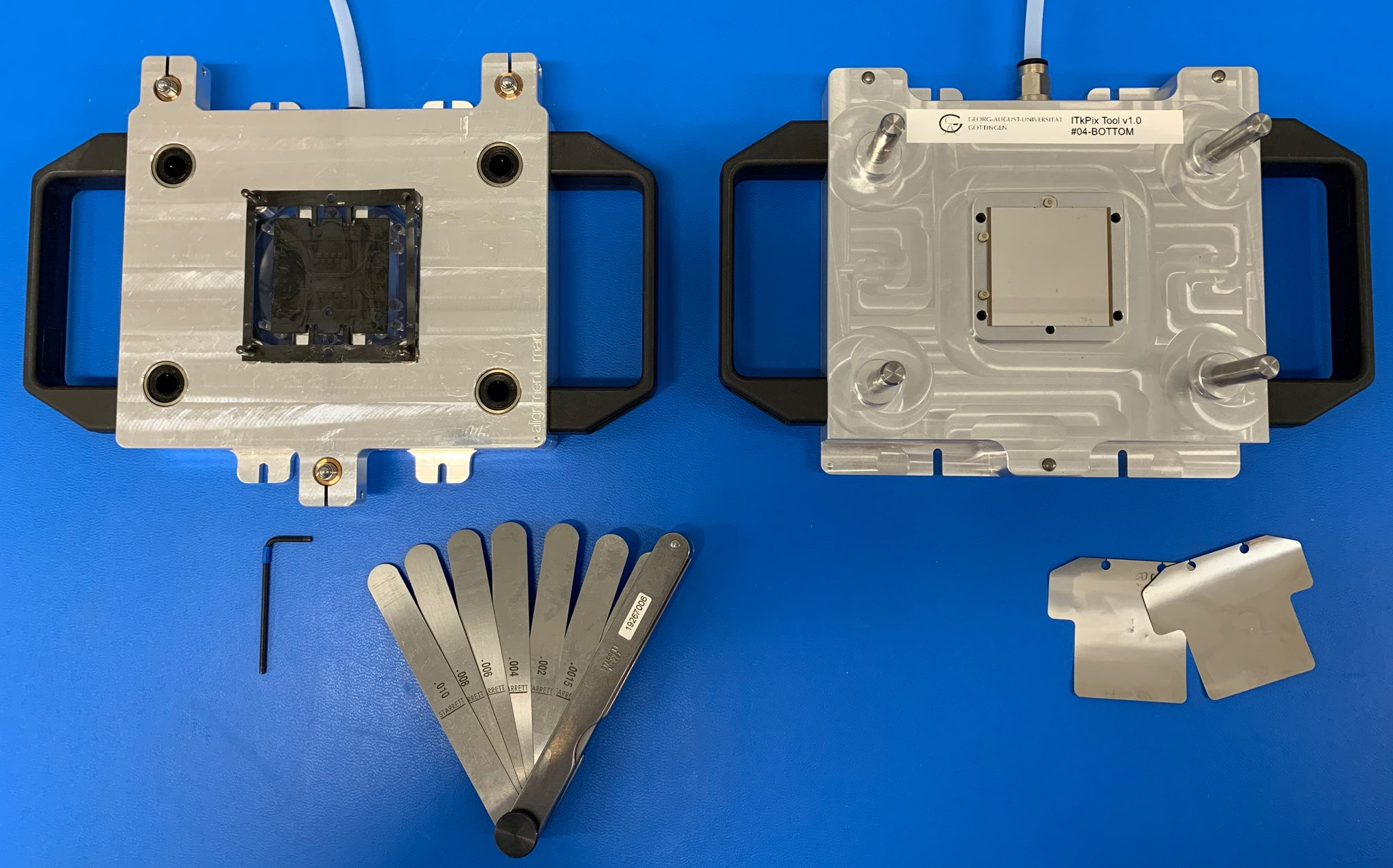}
\caption{Custom tools for pixel quad module assembly.}
\label{fig:assemblytooling}
\end{figure}

There are linear and round triplet modules in the innermost layer L0 for the barrel and rings, and quad modules for layers 1--4 making the largest part of the detector. About 10000 modules are going to be assembled for the ITk Pixel detector by hand during the next four years. Bare, hybridised sensor-frontend assemblies are glued onto module flexes using custom built tools, where glue is distributed by means of a laser-cut stencil and the thickness of the glue layer is controlled via precision shims as shown in Figure \ref{fig:assemblytooling}.

\begin{figure}[t]
\includegraphics[width=0.48\textwidth]{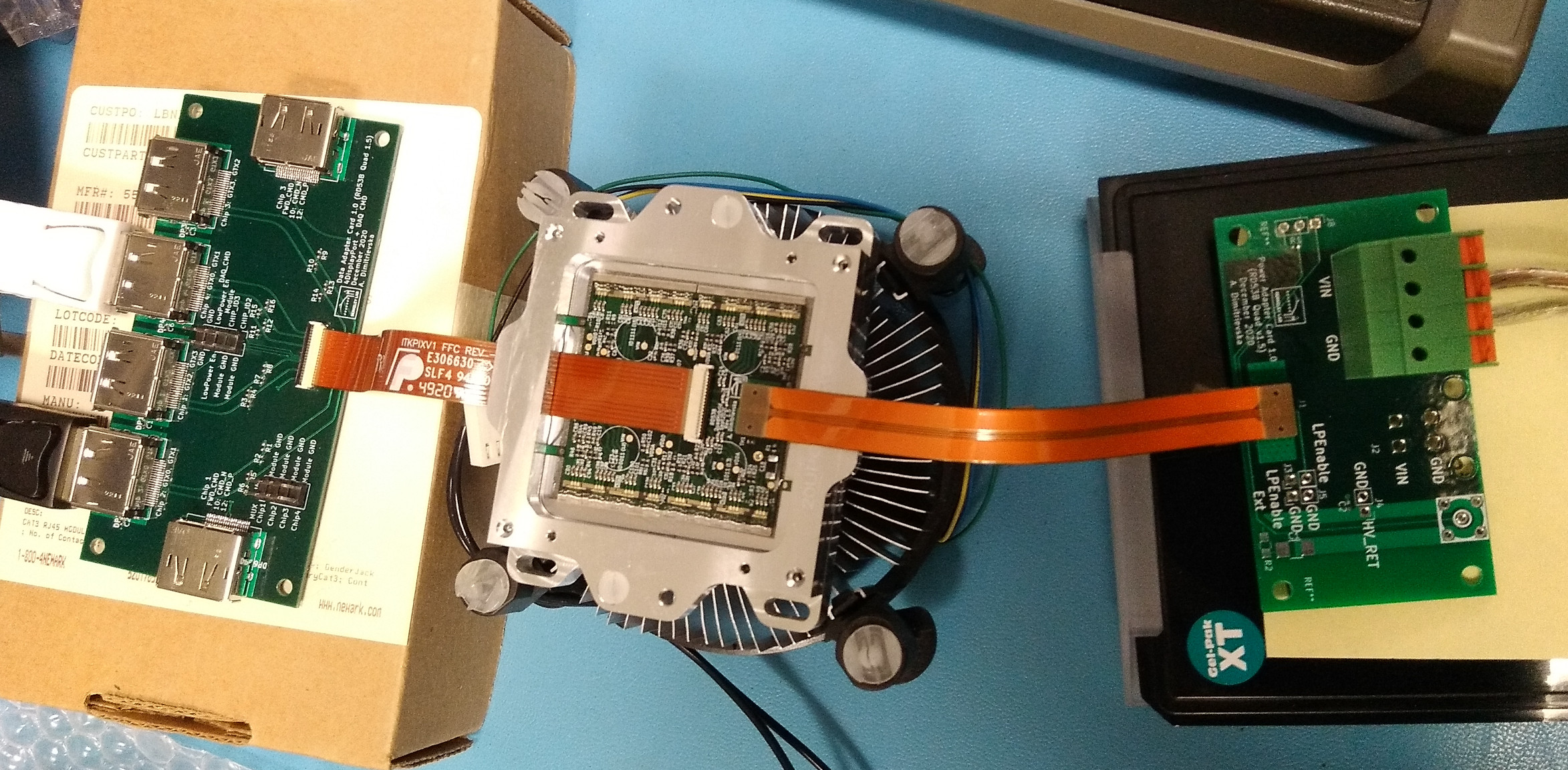}
\caption{One of the first ITkPixV1 digital quad module on a prototype flex PCB being tested.}
\label{fig:itkpixv1quad}
\end{figure}

Every assembled module has to pass a quality control procedure which involves visual inspection, metrology and electrical testing at different temperatures. Currently, the RD53A module programme is ongoing for institutes to set up infrastructure and gain experience in assembly, quality control and system test. Also first ITkPixV1 digital quad modules using prototype flexes are being tested with custom readout and power adapter boards as shown in Figure \ref{fig:itkpixv1quad}. These first tests verified the flex design but also allow access to probe multi-chip features like command forwarding, data aggregation etc.

\subsection{Serial Powering}
\label{sec:serialpowering}
\begin{figure}[t]
\includegraphics[width=0.4\textwidth]{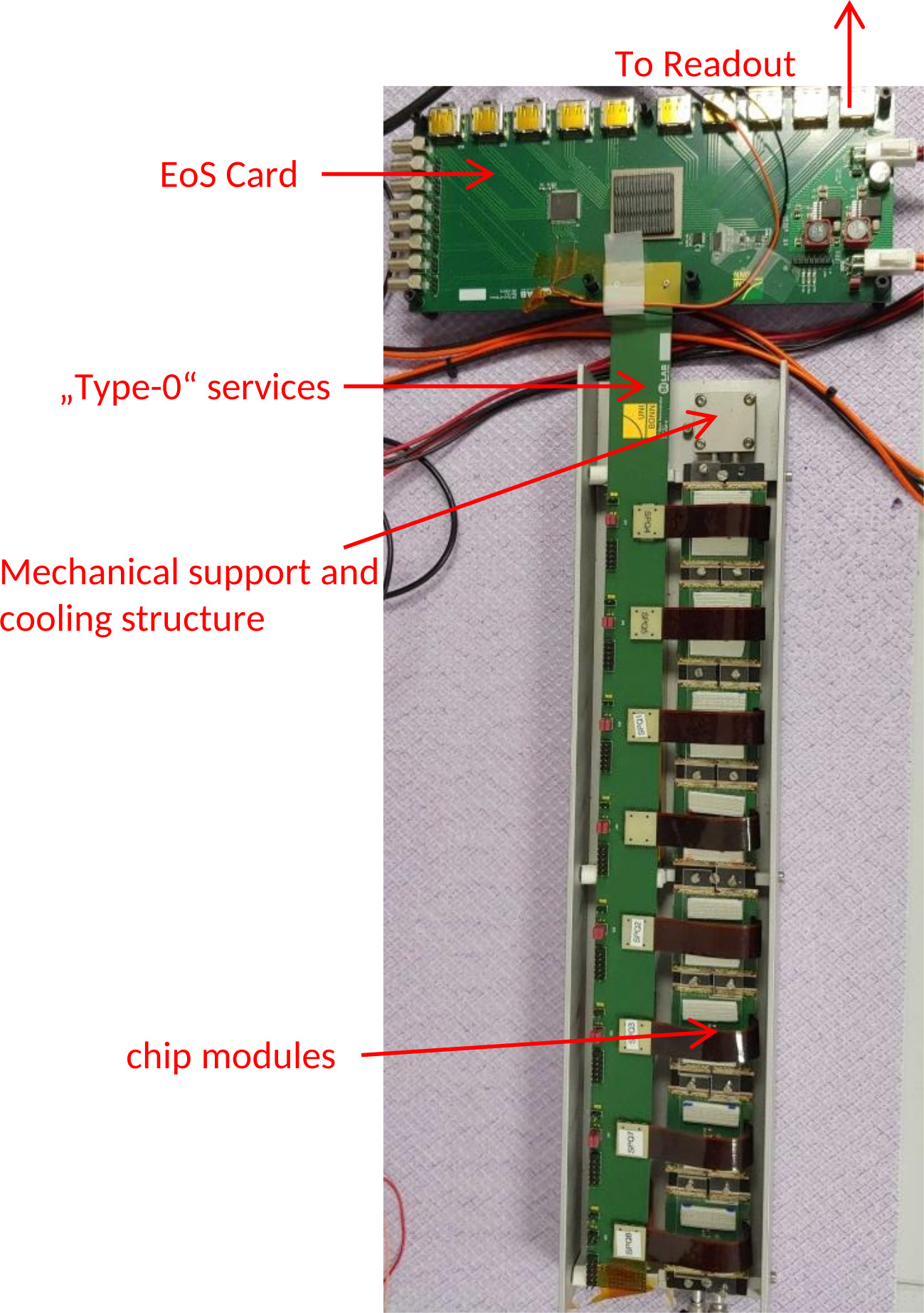}
\caption{Serial powering test with RD53A quad modules connected to pseudo Type-0 services and end-of-stave (EoS) card \cite{hinterkeuserIchep2020}.}
\label{fig:spchain}
\end{figure}

Readout chips on a triplet or quad module are powered in parallel using up to 8\,W of power. By powering up to 14 \cite{flickVCI2019} of those modules in series in a chain, the amount of required cables is reduced and thus material budget in the detector. This is made possible through shunt-LDO (SLDO) powering mode of the frontend chip by providing the chip with a constant and redundant power budget. Experience in operating serial powering chains was gained through demonstrators based on FEI4 modules \cite{fei4manual}. Studies on the first serial powering chain with RD53A quad modules is ongoing as shown in Figure \ref{fig:spchain} and chains with ITkPix modules are in preparation.

\begin{figure}[t]
\includegraphics[width=0.48\textwidth]{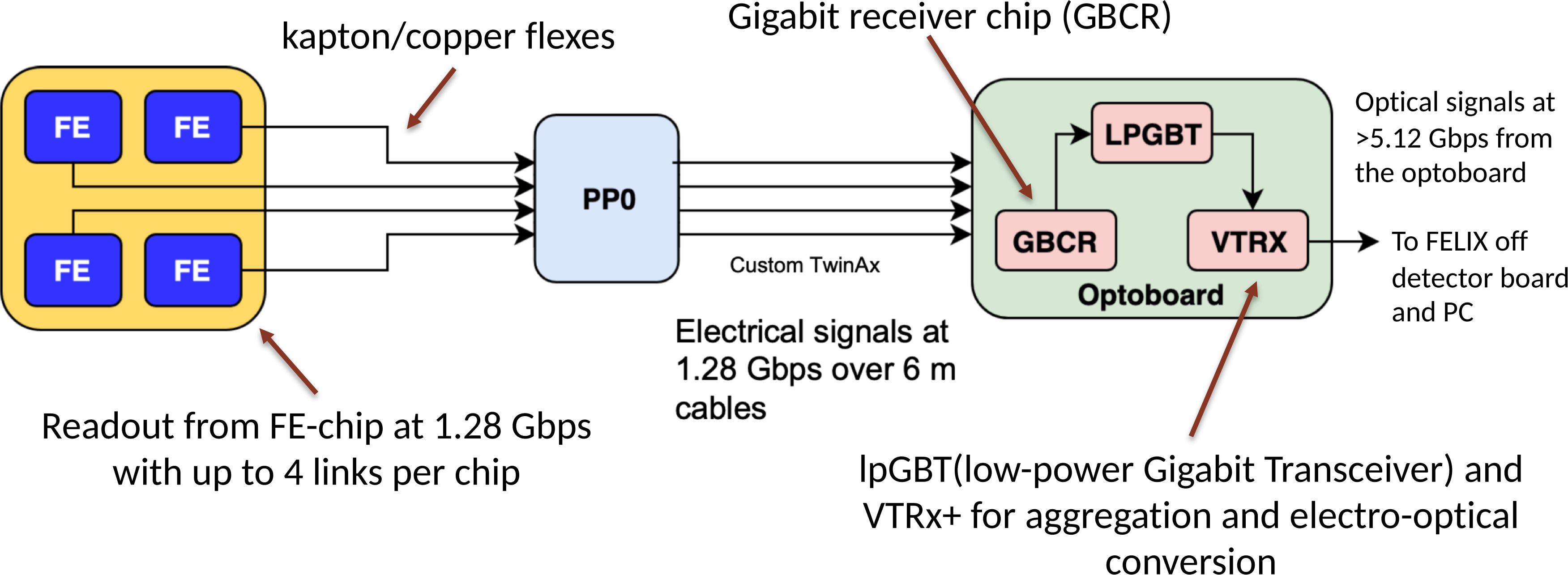}
\caption{Schematic of the data path \cite{gemmeVertex2020}.}
\label{fig:datatransmission}
\end{figure}

\subsection{Data Transmission and Services}
Each frontend chip is capable of transmitting data via 4 links with 1.28\,Gbit/s, resulting in up to 16 links on a quad module. The activity of a frontend chip depends on the hit rate and thus the position of the module in the detector. When full data rate is not required, the number of physical outgoing links per module can be reduced by aggregating data within the module through the frontend chips. The number of required data links can vary from 9 lines per triplet module in the innermost layer down to 1 line per quad module in L3. These links are shown in Figure \ref{fig:datatransmission} as kapton-copper flexes (Type-0 services) between module and patch panel 0 (PP0). From PP0 data is transmitted through custom Twinax cables to the optoboard \cite{optoboard}, where electrical signal is recovered by the Gigabit Receiver Chip \cite{gbcr} and is aggregated and converted to optical signal through the low-power Gigabit Transceiver (lpGBT \cite{lpgbt}) and VTRx \cite{vtrx} before being sent off detector to the readout.

\subsection{Mechanical Support}
\begin{figure}[t]
\centering
\includegraphics[width=0.48\textwidth]{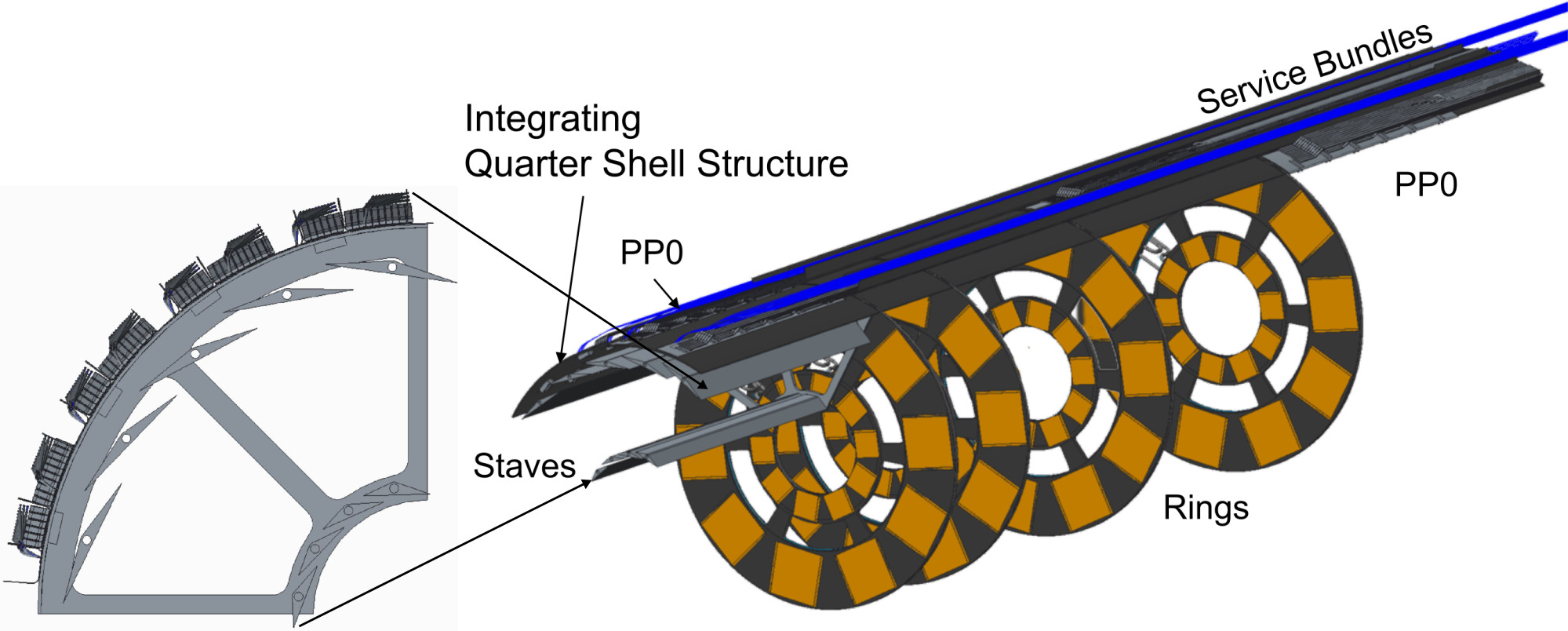}

(a)

\vspace{2mm}

\includegraphics[width=0.2\textwidth]{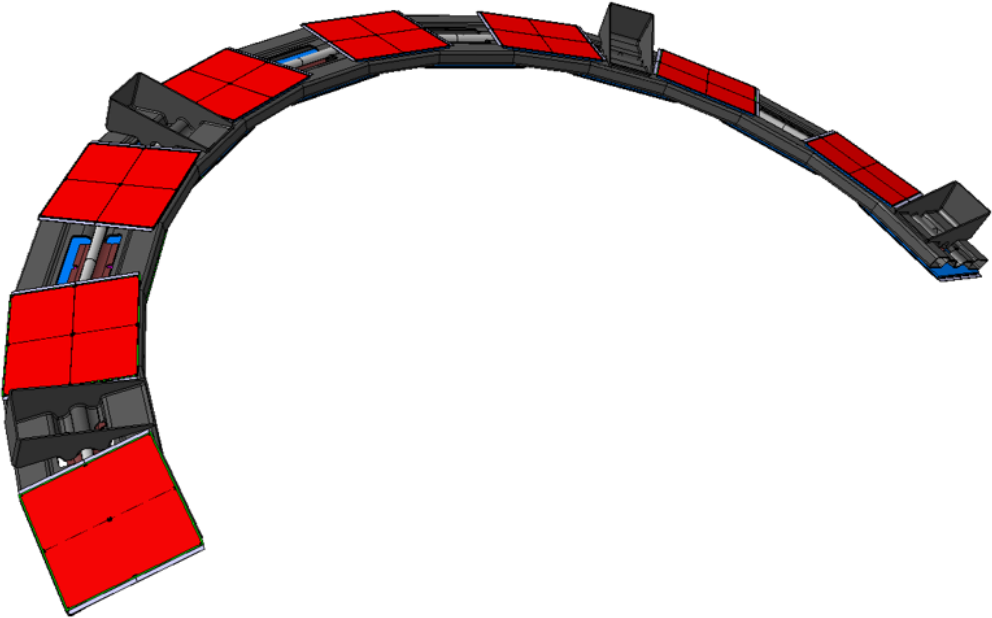}\includegraphics[width=0.3\textwidth]{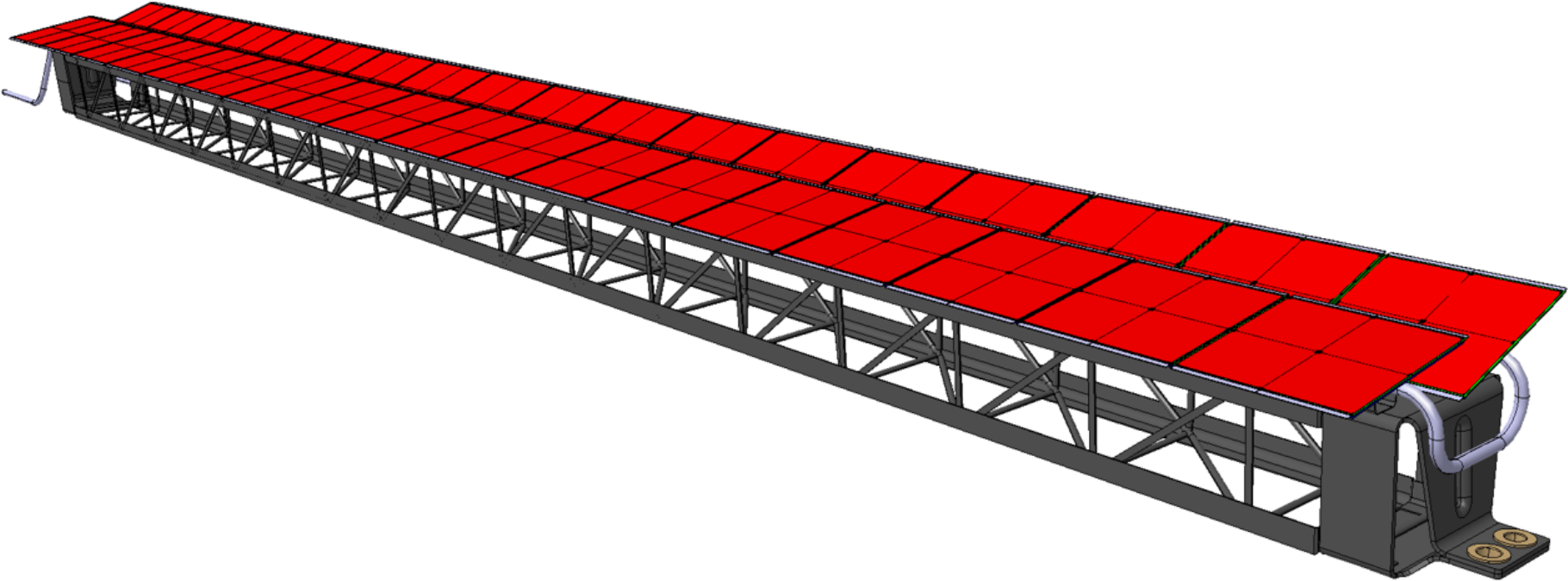}

(b)

\vspace{2mm}

\includegraphics[width=0.48\textwidth]{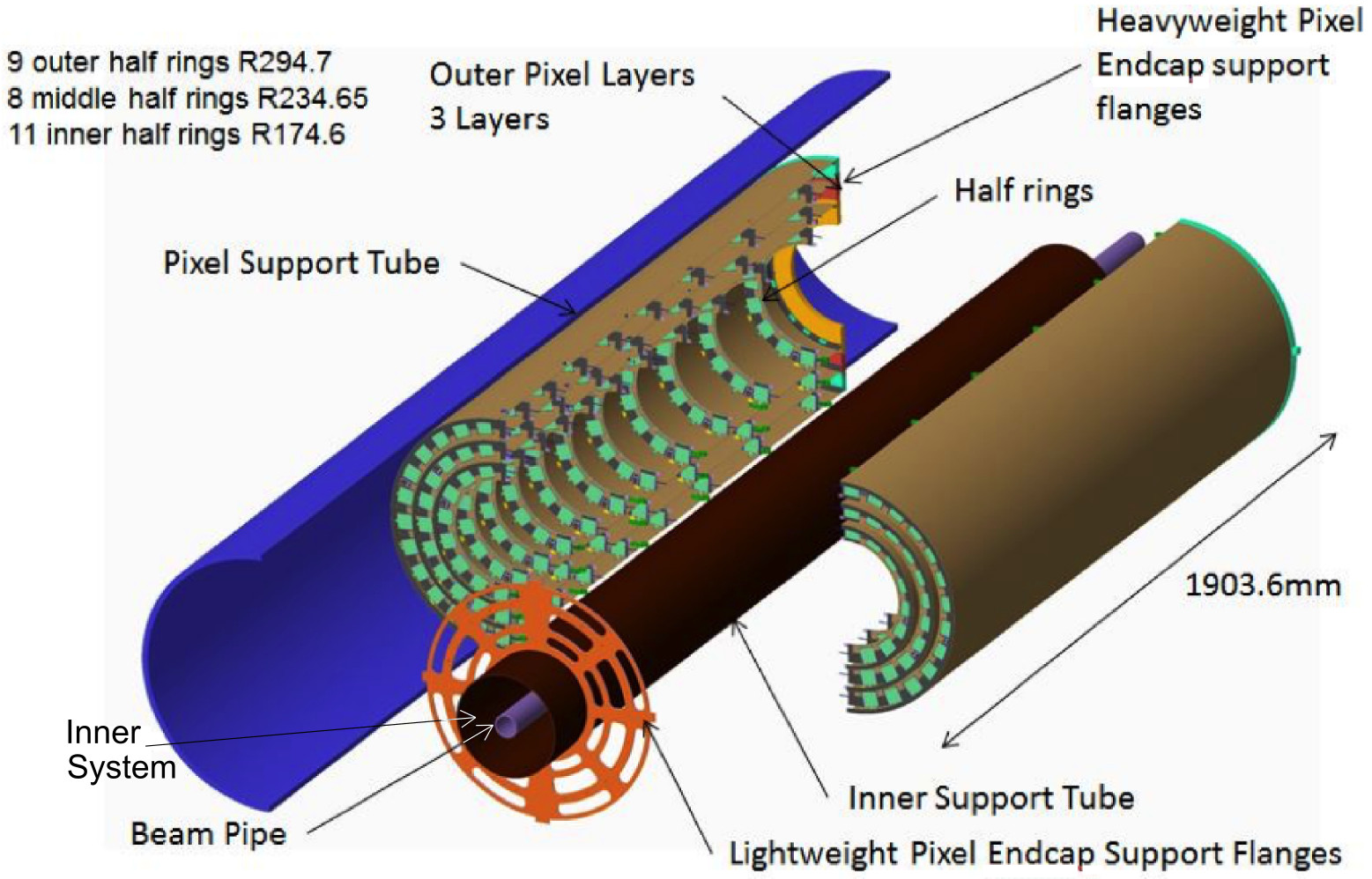}

(c)

\caption{Lightweight local support carbon fibre structures with pixel modules for the Inner System (a), Outer Barrel (b) and Outer Endcap (c) \cite{www:skuehnpixel2018, sutcliffeForum2019}.}
\label{fig:localsupport}
\end{figure}

In the detector, modules are attached onto and supported by lightweight carbon fibre composite structures, as shown in Figure \ref{fig:localsupport}. The Inner System uses a quarter shell concept for the barrel and endcap regions with different designs for the endcap rings (a). For the Outer System, the Barrel region (b) consists of longerons for the flat region and inclined half rings towards higher $|\eta|$, and Outer Endcap consists of modules that are attached onto half rings embedded in half shells (c).

\subsection{System Test}
\label{sec:systemtest}
\begin{figure*}
\centering
\includegraphics[height=41mm]{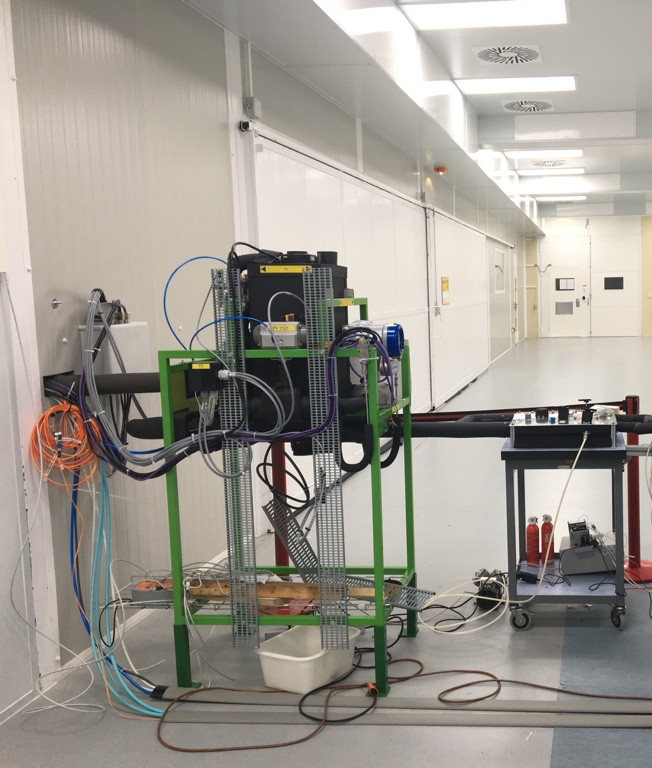}
\includegraphics[height=41mm]{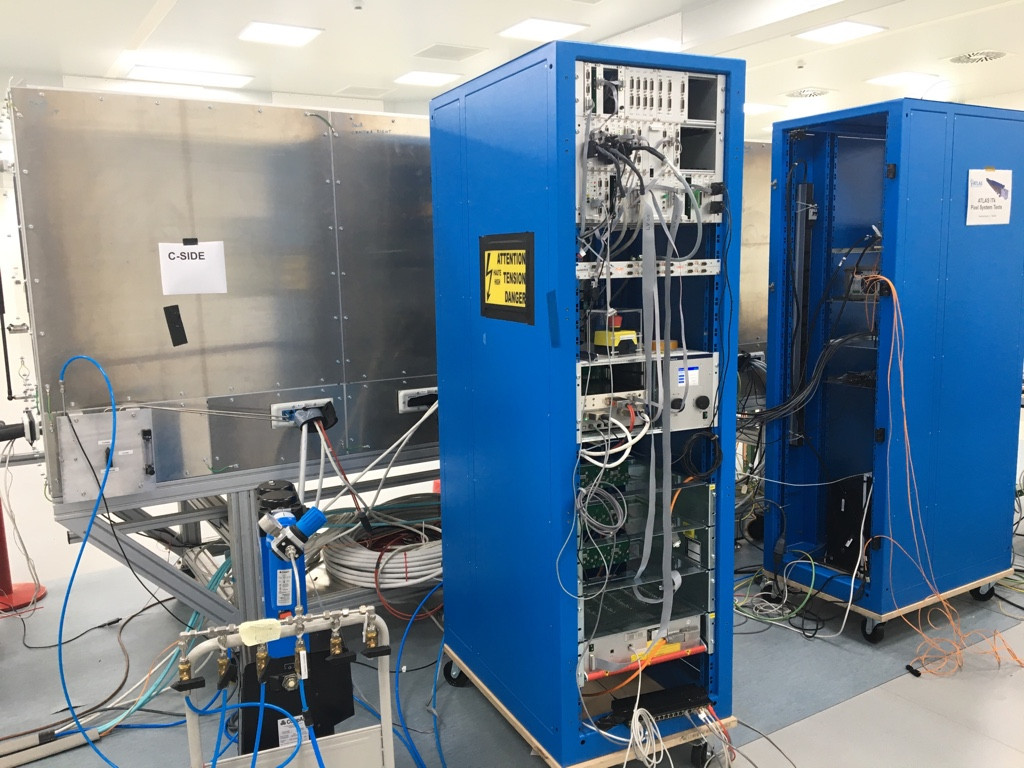}
\includegraphics[height=41mm]{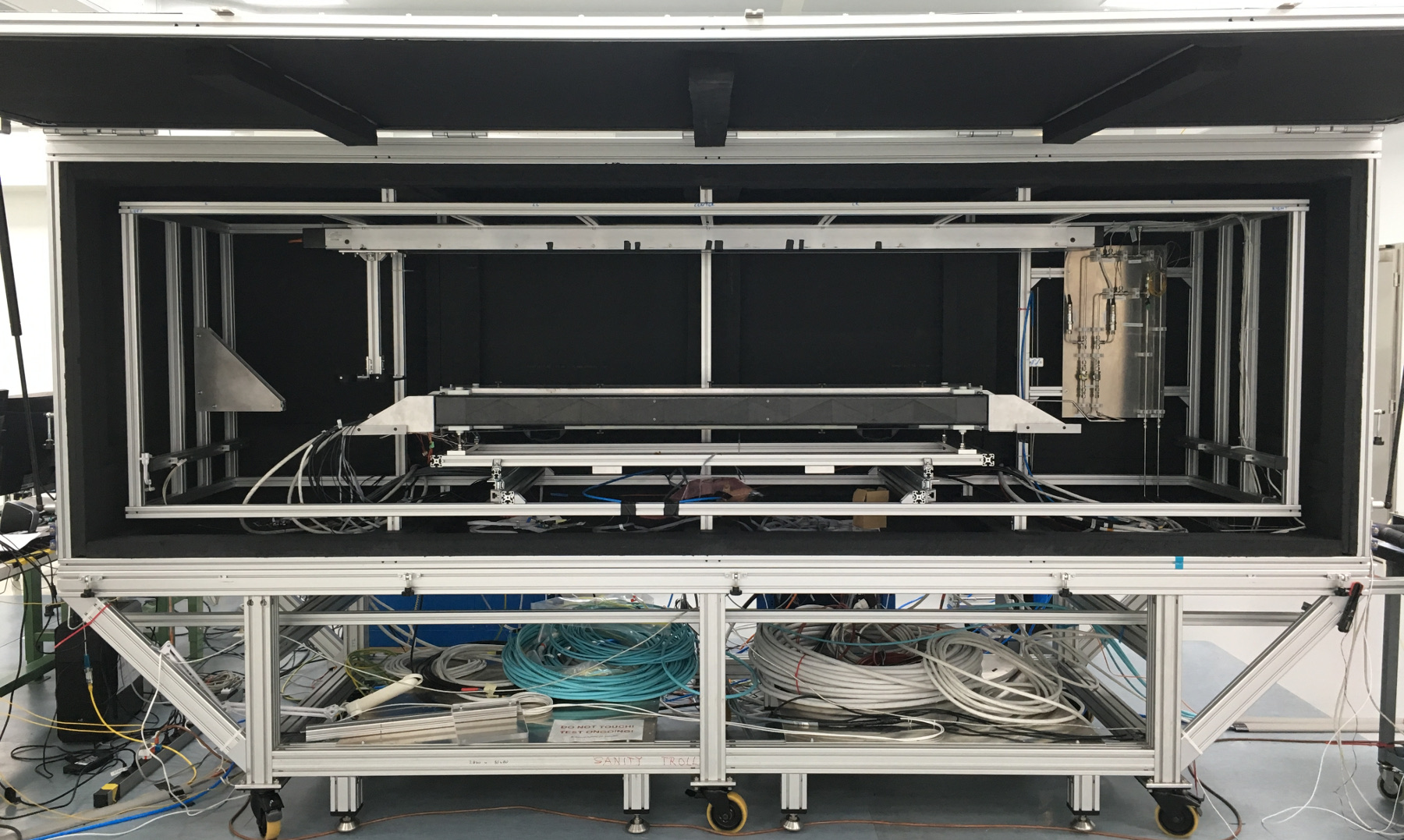}

\caption{System test infrastructure at CERN SR1. From left to right: CO$_2$ plant, dry air unit, racks with power supplies and interlock system, and light tight and thermally insulated testbox \cite{mengTwepp2019}.}
\label{fig:sr1}
\end{figure*}

System test with realistic module staves and infrastructure is ongoing for all subsystems of the pixel detector, using all components introduced above. Figure \ref{fig:sr1} shows an example of the system test infrastructure for Outer Barrel at CERN, consisting of a CO$_2$ cooling plant, a dry air unit, racks with power supplies, detector control system (DCS), an interlock matrix, and a light-tight and thermally insulated testbox with a source stage.

\begin{figure*}
\centering

\hspace{0.03\textwidth}\includegraphics[width=0.82\textwidth]{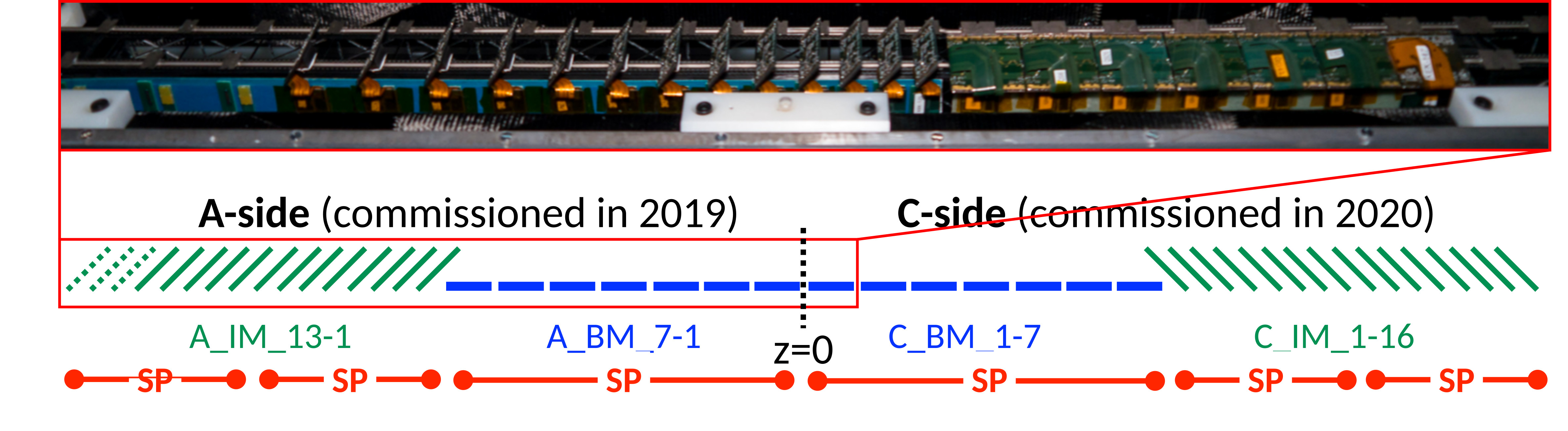}\hspace{0.05\textwidth}

\includegraphics[width=0.85\textwidth]{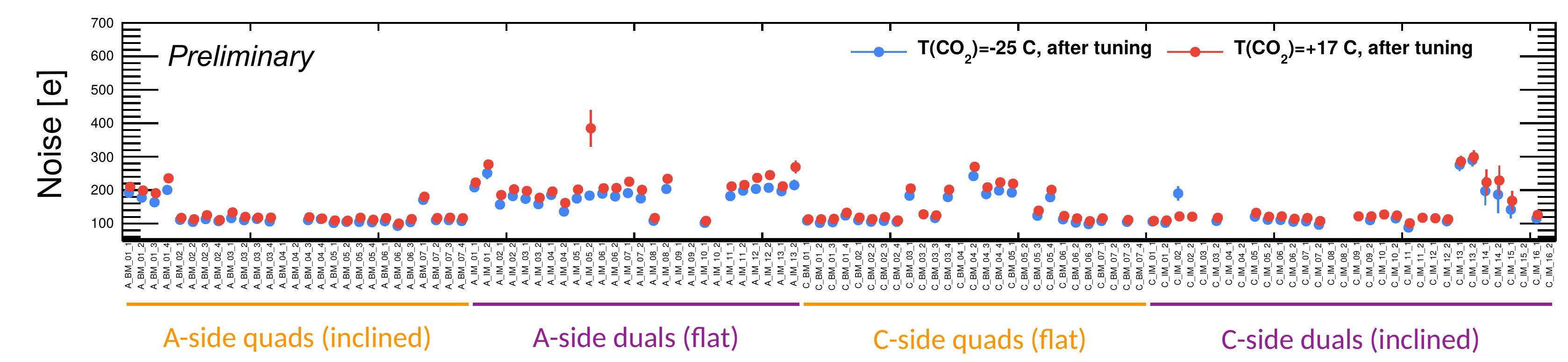}\hspace{0.05\textwidth}

\hspace{0.03\textwidth}\includegraphics[width=0.84\textwidth]{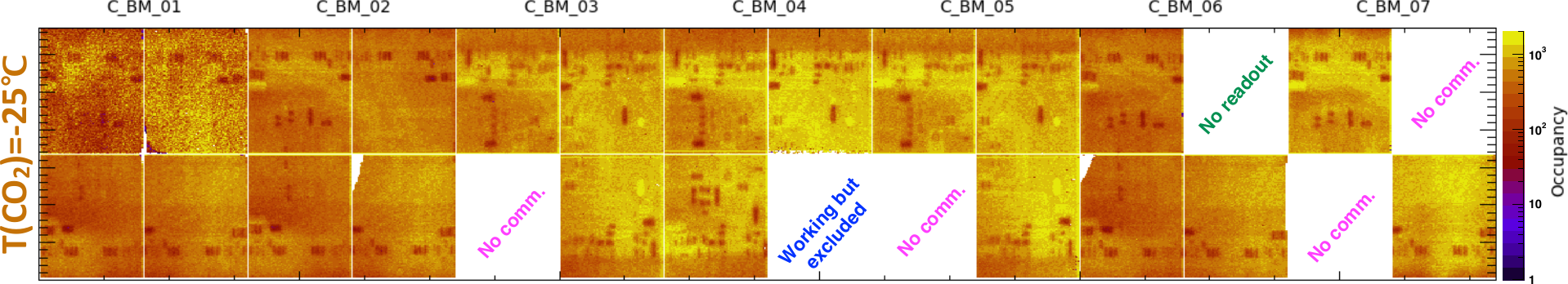}

\caption{Results from the Outer Barrel demonstrator using prototype FEI4 modules. Top: layout of the modules. Middle: noise level of the modules measured warm and cold with CO$_2$ cooling at $17\,^\circ$C and $-25^\circ$C, respectively. Bottom: source scan of the C-side quad modules at $-25^\circ$C using $^{90}$Sr sources \cite{zambitoTrento2021}.}
\label{fig:demonstrator}
\end{figure*}

The demonstrator is a full-size, 1.6\,m long prototype stave that is loaded with FEI4-based modules \cite{zambitoTrento2021}. The module arrangement is indicated in Figure \ref{fig:demonstrator} top: with 14 flat quad modules in the centre, 7 on each A-side and C-side, and 13 and 16 inclined dual modules towards high $|z|$ on the A-side and C-side, respectively. The modules are connected in 6 serial powering chains. The middle plot shows a noise level of all modules, measured warm and cold with CO$_2$ cooling at $17\,^\circ$C and $-25^\circ$C, respectively. As expected the noise level decreases with lower temperature. A source scan of the C-side quad modules performed at $-25^\circ$C is shown in the bottom plot by running two $^{90}$Sr sources on the stage over all modules. The quad modules on the C-side were fully functional on the module level. After integration onto the stave some frontend chips lost communication which is mostly due to issues related to cables, except for the chip on BM\_03 which shows a perfect source scan when warm \cite{zambitoTrento2021}. The path leading to the cold source scan of the demonstrator provided valuable experience and demonstrated a working constellation of system infrastructure.


\section{Summary}
This paper has given a brief overview of the status of the pixel system of ATLAS Inner Tracker upgrade. Module components, sensor, frontend chip and module flex are being finalised. Module institutes are set up and being qualified for assembly and testing so the RD53A module programme can progress with full speed. With these modules, the gained experience with serial powering tests and fully functional FEI4 prototype systems preparing for the next generation demonstrators with the first RD53A modules to validate every aspect of the ITk Pixel detector development.


\bibliographystyle{unsrt}
\bibliography{20210318_LCWS_ITkPixel_lmeng}

%
%
%


\end{document}